# Photo luminescence of Cooper pairs in a naturally occurring heretostructure $K_{0.75}Fe_{1.75}Se_2$

A. Ignatov[1], R.H. Yuan[2], and N.L. Wang[2]

Combining superconductor and semiconductors in nanostructure junctions was a challenging technological problem that attracted attention for long time [van Wees]. The radiative recombination of Cooper pairs was demonstrated, using a Nb/n-InGaAs/p-InP heterostructure [EL_Hayashi, PL_Hayashi], called Cooper pair LED [PL_Suemune, R_Suemune]. It has been suggested that the junction could produce entangled photon pairs [Benson, Gywat] needed for quantum information processing and communication.

Here we demonstrate an enhanced radiative recombination of electron Cooper pairs in inhomogeneous $K_{0.75}Fe_{1.75}Se_2$ (KFS) subjected to laser light upon cooling below superconducting transition temperature $T_c \sim$ 28 K. The observation of this phenomena is possible due to fulfillment of the following three conditions: (1) Phase separation in superconducting KFS crystals is realized via naturally occurring heterostructure [mic_Charnukha, TEM_Wang]; (2) Partial Fe-vacancy ordered *n*-type semiconducting regions, sandwiched between Fe-vacancy free SC and vacancy-ordering AFM structure, forming active layers. The electronic structure of those active layer is tuned to induce Cooper pairs by the proximity effect [de Gennes] and/or to accept Cooper pair tunneling from the SC phase. In the active layer, the electron Cooper pair can radiatively recombine with two *p*-type holes [Asano] produced by laser photoexcitation; (3) In KFS nature provides at least $10^9$ SC/n-type semiconducting/AFM insulating junctions per $cm^3$.

For laser excitation energy of 1.92 eV ($\lambda$=647.1 nm) and power 0.68 mW focused to the spot of 40x80 $\mu m^2$ the estimated internal quantum efficiency of the natural heterostructure at 10 K in the luminescence range of 700 to 1300 nm is about 30 % and is likely limited by availability of the *p*-type holes.

[1] Department of Physics and Astronomy, Rutgers The State University of New Jersey, Piscataway, New Jersey 08854, USA. [2] Beijing National Laboratory for Condensed Matter Physics, Institute of Physics, Chinese Academy of Science, Beijing 100190, China.

Correspondence and request for materials should be addressed to A. I. (aignatov@physics.rutgers.edu)

# Introduction

(1) Combining superconductor and semiconductors in nanostructure junctions was a difficult technological problem that attracted attention for long time [van Wees]. Such junction is potential (1) bright light source [Hanamura], and (2) entangled photon pairs (EPPs) source, needed for quantum information processing and communication [Benson, Gywat]. The main technological problem was assembling heterostructures out of two classes of materials having essentially different crystallographic and electronic structures. Recent progress has been made by Suemune's group [EL_Hayashi, PL_Suemune, R_Suemune]. Nb/n-InGaAs/p-InP heterostructure based on the state-of-the-art InGaAs technology were build and demonstrated enhanced radiation recombination (high light output), but not EPPs generation.
In $K_{0.75}Fe_{1.75}Se_2$ (KFS) an intrinsically inhomogeneous material [mic_Charnukha, TEM_Wang, IR_Yuan, STM_Li, IR_Homes, tr1_Ryan, NQR_Texier, ARPES_Chen, ND_Zhao] belonging to $A_yFe_{2-x}Se_2$ (A= Rb, K, Cs) family of Fe-chalcogenides, nature offers an elegant solution of the problem: Superconducting, Fe-vacancy free domains ($K_{y1}Fe_2Se_2$) are epitaxially bonded to semiconducting layers ($K_{y2}Fe_{2-x}Se_2$) characterized by gradient of Fe-vacancies. Phase separation in KFS is realized via naturally occurring heterostructure [mic_Charnukha, TEM_Wang].

(2) Here we demonstrate an enhanced radiative recombination of electron Copper pairs in $K_{0.75}Fe_{1.75}S_2$ subjected to laser light upon cooling below superconducting transition temperature $T_c$~28K. In agreement with inhomogeneous nature of the sample, the luminescence output was found essentially spot-dependent. However, the similar slightly asymmetric Gaussian lineshape (centered at ~930±20 nm, FWHM ~260 nm) and the threshold-like enhancement of the luminescence output below $T_c$~ 28 K were observed regardless of the spot position. Based on the luminescence output as a function of laser power the enhanced radiative recombination was found to be limited by availability of quasi-equilibrium *p*-type holes produced by laser pumping, down to the lowest measured temperature of 6 K. Using a simple model, assuming that the active layer is located inside the effective medium, we evaluated the internal quantum efficiency (IQE) of the laser-pumped KFS heterostructure, defined as η= $N_{out}(λ=930nm)/N_{in}(λ=647nm)$ (number of photons "out" to number of photons "in"). The IQE is found to be close to 30%, rivaling or exceeding the IQE of the best human-made heterostructures [PL_Suemune, R_Suemune].

(3) For the SC ($K_{y1}Fe_2Se_2$) / n-type semiconducting ($K_{y2}Fe_{2-x}Se_2$) junction not only are the crystal structures well matched, but the electronic structure of active layer is tuned to induce Cooper pairs by proximity effect [de Gennes] and/or to accept Cooper pair tunneling from the SC phase. The semiconducting phase admits diffused *p*-type holes generated in Fe-deficient regions by laser light. In the active regions of KFS, an ensemble of electron Cooper pairs radiatively recombines with *p*-type holes (every Cooper pair annihilating two *p*-type holes), giving rise to partially coherent luminescence.

## Results

(1) Luminescence and Raman scattering are two constituents of the secondary optical emission. Fig. 1 provides an overview of the secondary emission of $K_{0.75}Fe_{1.75}Se_2$ (KFS) for laser excitations of λ=647 (red), 568 (orange), and 413 nm (purple). In the energy range accessed in this paper, there are two luminescence bands: a narrow one at ~$1.1\times10^4$ and broad band at ~$2.1\times10^4$ cm$^{-1}$ (1 eV = 8065.7 cm$^{-1}$) The low-energy luminescence suddenly "switches on", gaining at least one order of magnitude as temperature drops from ~40 to 10 K crossing $T_c$~ 28 K. The high-energy luminescence does not exhibit the threshold behavior and will not be discussed further here.

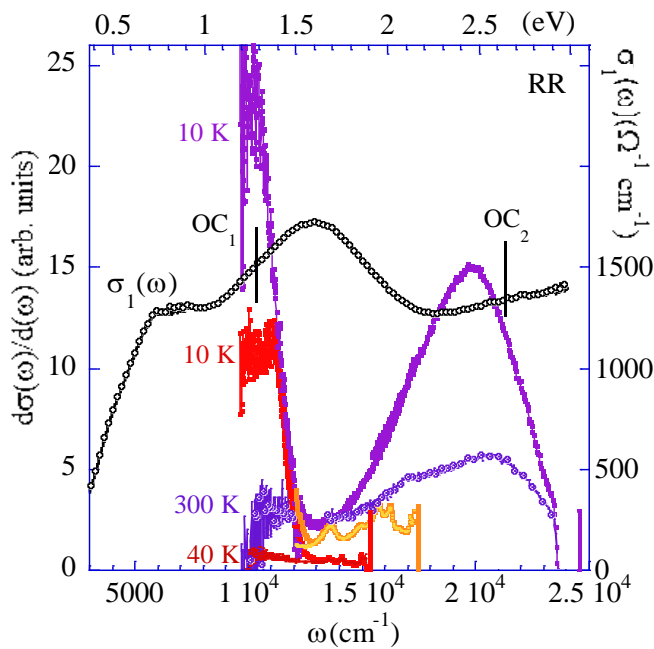

**Figure 1. Secondary emission of $K_{0.75}Fe_{1.75}Se_2$ in the energy range of ω=(1.-2.5)x10$^4$ cm$^{-1}$ (1.2-3.0 eV) at T=10 (solid lines), 40 K (dashed), and 300 K (open circles).** Three different laser excitations of λ= 647, 568, and 413 nm (vertical bars at 1.92, 2.18, and 3.01 eV) are shown by red, orange, and purple. The experimental optical conductivity spectra σ$_1$(ω) of the same sample at 300 K from Ref. [IR_Yuan] (black open dots) is given for comparison. Since the band positions do not depend on the excitation energy the bands are luminescent in nature. The bands appear near the bottoms of corresponding optical conductivity bands (labeled OC$_1$, and OC$_2$). Sharp data cut-off at ~1.23 eV and high-frequency modulation at low energies are both due to Si-based CCD detector limitations, addressed by use of InGaAs linear detector, see to Fig. 2.

(2) Repeated measurements were undertaken for λ=647 nm excitation in order to (a) understand how luminescence output depends on local spot position on the intrinsically inhomogeneous sample; (b) to extend the spectral range down to 0.95 eV by employing InGaAs detector; and (c) to provide accurate temperature and power dependences of the low-$T$ luminescence and estimate internal quantum efficiency of the laser-pumped KFS heterostructure.

(3) The luminescence output was, indeed, found to be essentially spot dependent. Five different spot positions were explored. Spectral radiance, $SR^0_{exp}(\theta=0°,\lambda)$ was measured in the direction perpendicular to the surface (along the crystallographic *c*-axis). Results for two spots characterized by highest(lowest) photo luminescence are shown in Fig. 2 a(b). Despite the 200x amplitudes difference the following three observations holds true for all measured spots: (a) The emitter lineshapes are well approximated by slightly asymmetric Gaussians centered at 930±20 nm and FWHM ~ 260 nm. (b) The spectral radiance exhibits threshold-like enhancement below $T_c$~ 28 K. (c) The luminescence output tends to saturate somewhat below 12 K.

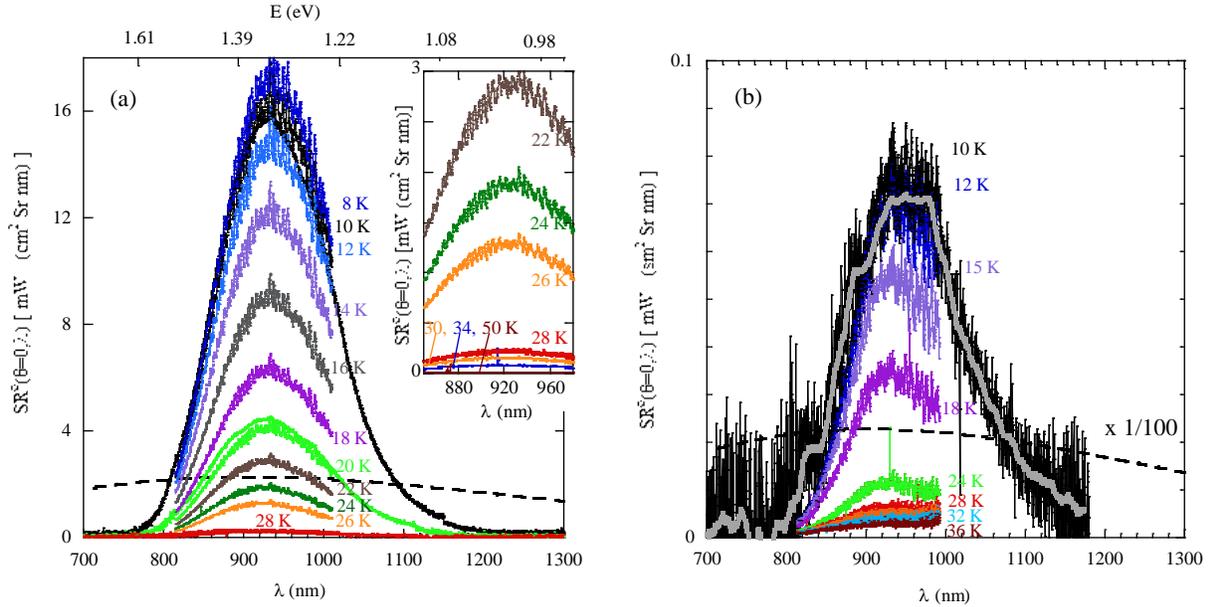

**Figure 2**. **Temperature dependence of the spectral radiance along the normal to the surface** (band at ~1.33 eV in Fig. 1) from the best (a) and the worst (b) luminescence-generating spots (out of total 5 measured spots) for laser power of 0.68 mW focused to the spot ~40x80 µm². The radiances obtained from Si CCD [fine solid in the range of 820-1000 nm at 8-28 K, and 22-50 K in insert to FIG. 2(a)] are compared with the radiances inferred from InGaAs measurements over the range of 655 to 1300 nm at 10, 20, and 28 K and with the radiance of the black body calibration source (dashed line). "Switching on" the low-*T* luminescence at 26-28 K is apparent from insert to Fig. 2(a), refer also to Fig. 3. For 2(b) the luminance increments at 26-28 K, followed by larger jump at 18-24 K. For 2(a), position and FWHM of the PL peak do not change within 30 nm over the range of 8-28 K, implying that inhomogeneous broadening dominates the width. A radiance, $R^o_{exp}(\theta=0°)$, is the integrated spectral radiance $SR^o_{exp}(\theta=0°, \lambda)$ over the λ-range of 700-1300 nm. For 2(a) at T=10 K, $R^o_{exp}(\theta=0°)$ = 2.86 x10³ mW/(cm² Sr).

(4) Amplitude of radiance as function of temperature for the best spot position is shown in FIG. 3. On the cooling down, small progressive increase of the "background" luminescence (50 → 32 K), followed by sharp upturn (28 → 14 K) and saturation regime (T < 10 K). The amplitude raises ~ 26 times as temperature decrease from 30 to 10 K. Over the range of 14-28 K, the experimental data can be satisfactory fitted by $C_1(T_c/T-1)$, were $C_1$= 11.46 mW/(cm² Sr nm) and $T_c$= 28 K (solid green line). Assuming $\Delta(T) \sim \alpha(1-T/T_c)^{1/2}$ from BCS or Ginzburg-Landau theory at T→$T_c$ the experimental data follows closely the $\Delta^2(T)/T$ functional dependence at 0.6<T/$T_c$<1.0. The saturation of luminescence output can be fitted by $C_2 \Delta^*(T)$ with somewhat reduced $T^*_c \sim 0.6\, T_c$ (solid grey line) at $0.21 < T/T_c < 0.5$. The low-$T$ saturation is observed at 0.68 and as low as 0.04 mW of laser power used.

(5) The spectral radiance, not normalized to power, is a monotonically raising function of laser power. However, for all three spots where the luminescence vs. power dependences were collected, the amplitude of spectral radiance normalized to power, $SR^0(\theta=0°, \lambda=930\text{nm}, P)/P$ curves are scalable: being multiplied by constants they all full on a universal curve. Choosing the scaling constants to have $SR^0(P)/P \to 1.0$ at low powers we obtained a set of curves shown in Fig. 4. Luminescence is linearly proportional to power only at $P < P_c \sim 0.12$ mW (blue area). At higher powers a new relaxation channel opens up: photoexcited $p$-type holes tend to recombine with photoexcited electrons locally, rather than to diffuse and take part in the enhanced radiative recombination that will be discussed later. Good matching of data collected at 6 and 45 K (well below and above $T_c$) implies that the $e$-$h$ relaxation takes place mostly in the insulating phase.

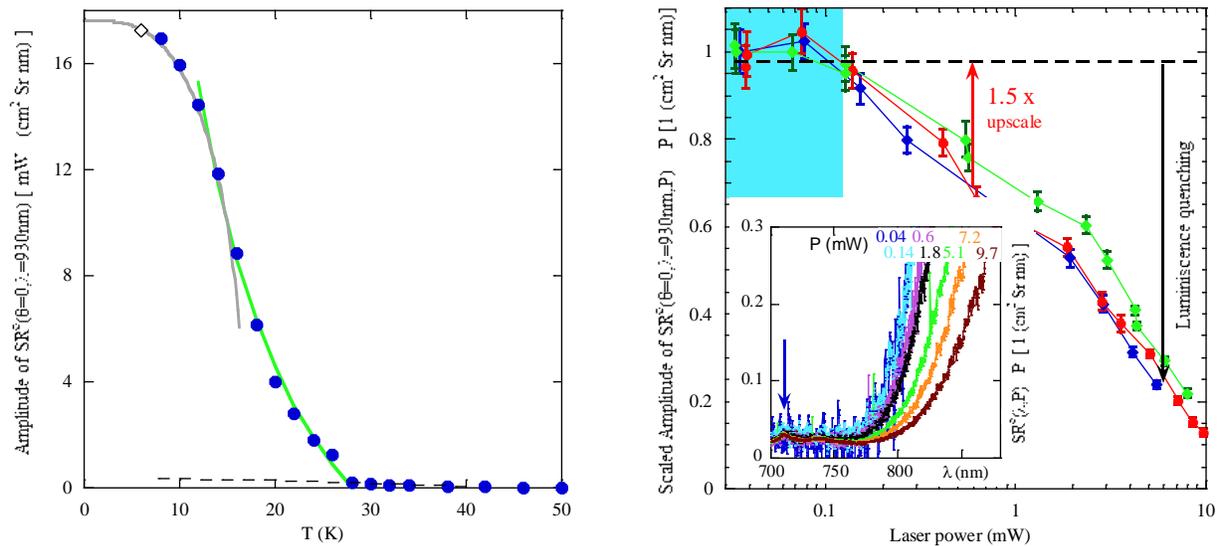

**Figure 3 (left). Amplitude of the spectral radiance as function of temperature**, based on experimental data from Si CCD, FIG 2(a). The 6K data point (open rhomb) is added from luminescence vs. power dependence. The error bars are less than the symbol sizes. Green(grey) solid line is a fit to the experimental data over the range of 14-28 (6-14K). The "background" luminescence existing in the whole measured $T$-range (extrapolated below 28K) is shown by dashed line.

**Figure 4 (right). Luminescence quenching as a function of laser power (logarithmic scale) delivered to the sample for three different spot positions.** Blue dots: insulating spot S3, the $SR^o(T, \lambda, P=0.68\text{mW})$ data are shown in Fig. 2(a), sample was maintained at 6 K; Red dots: metallic spot M2, the $SR^o(T, \lambda, P=0.68\text{mW})$ are shown in FIG 2(b), sample at 6 K; Green dots: insulating spot S2, no data shown, sample at 45 K (well above the $T_c$). Inset shows the $SR^o(\theta=0°, \lambda, P) / P$ for the metallic spot M2 (red dots in Fig. 4). Note that amplitude of Raman features at ~720 nm (blue arrow) is directly proportional to the power over the range 0.04-9.7 mW (spanning more than 2 orders of magnitude) but the Cooper pairs luminescence is linearly proportional to power only up to $P_c < 0.12$ mW (blue area, in FIG. 4). At higher powers, the luminescence is rapidly quenched due to an extra relaxation channel for the photo excited $p$-type holes. Red arrow indicates up-scaling factor (~ 1.5) implying that if the internal quantum efficiency were measured at $P < 0.12$ mW, it would be 1.5 times higher than measured at $P = 0.68$ mW.

**Discussion**

(1) To understand an origin of the observed luminescence we shall briefly review a microscopic structure of KFS. Recently Charnukha *et al* [mic_Charnukha] studied phase separation in the single-crystalline antiferromagnetic superconductor $Rb_2Fe_4Se_5$ using a scattering-type scanning near-field optical microscopy. They demonstrated that the AFM and SC phases segregate into nanometer-thick layers. Based on their results one can deduce three conclusions regarding morphology of metallic/SC regions: (a) The regions are perpendicular to the Fe-Se planes; (b) They are ~10 μm long, up to ~ 10 μm high (out-of-plane dimension), and less than 1 μm wide; (c) The regions are perpendicular to each other, terminated by running into perpendicular neighbor, and occur at 45° with respect to the in-plane crystallographic axes.

The optical microscopy results agree with high-resolution TEM imaging [TEM_Wang]. At least within the area shown in Fig. 8 in Ref. [TEM_Wang] an **epitaxial-like arrangements** between ordered AFM and SC phases is observed. Homes *et. al*. [IR_Homes] modeled in-plane optical properties of inhomogeneous $K_{0.8}Fe_{2-y}Se_2$ with $T_c$= 31 K in the normal state using the Bruggeman effective medium approximation (EMA) for metallic inclusions in an insulating AFM matrix. The volume fraction of the metallic phase was estimated to be ~ 10%. EMA can only successfully fit the IR reflectivity data if the inclusions are highly distorted and overlap to some degree, suggesting a filamentary network of conducting regions joined through weak links, in agreement with results of the near-field microscopy. Nanoscale phase separation between superconducting and ordered AFM phases was also identified by the observation of sharp Josephson plasma edge in infrared reflectance spectrum at low frequency below $T_c$ [IR_Yuan].

Since smallest linear dimension of metallic/SC domain is much larger than SC coherent length (ξ~ 2 nm) [tr1_Ryan], the typical individual brick of $V$=10x10x1 μm$^3$ exhibits bulk superconductivity on $T_c$. Assuming the volume fraction of the metallic phase in KFS matrix to be *f*~ 0.1 (10%) [IR_Homes, tr1_Ryan] there are $N_d$=f/$V$~ $10^9$ superconducting bricks/domains in cm$^3$. Since near-field IR imaging [mic_Charnukha] treats doped $K_yFe_{2-x}Se_2$ as being metallic for much high *x* than NQR [NQR_Texier], the above is the low-side estimate for the number of SC domains per cm$^3$.

(2) Three-phase models of phase separation in the $K_yFe_{2-x}Se_2$ are clearly getting traction [Supplementary Note #1]. Partly Fe-vacancy ordered semiconducting phase serve as a buffer between SC and AFM phases where the gradient in the Fe-vacancy concentrations is leveled out. The *n*-type semiconducting regions are sandwiched between Fe-vacancy free SC and vacancy-ordering AFM structures. Based on the above estimate for concentration of SC domains, we ought to conclude that **in KFS nature created at least $10^9$ SC/n-type semiconducting/AFM insulating heterostructures per cm$^3$**. Thin layers adjoined to the "true" metallic/SC phase, form active layers, where conditions for radiative recombination of Cooper pairs are optimal, as discussed in the following section.

(3) The heterostructure is self-assembled in a way that SC/n-type semiconducting layers are coupled in plane, along **a**+**b** or **a**-**b** crystallographic directions. In KFS superconductors the coherent length in the *ab*-plane is 2-4 times longer than along the *c*-axis. Thus, the way the heterostructure naturally grown facilitates propagation of the Cooper pairs from SC to the *n*-type semiconductor active layers.

**The mechanism of photo luminescence in KFS.**

(1) Due to Fe *d*- and Se *p*- hybridization, a *p*-type band topping at ~ 1.3 eV below $E_f$ is formed within broader and higher density *d*-band [LDA_Nekrasov], Fig. 5. Laser excitation (green arrows) couples occupied *p*- to empty *d*- via dipole transition for all allowed states in the BZ. Due to frequent *e–h* collisions photo-excited hole submerges at the top of the *p*- band, in a valley near the Γ-point[LDA_Nekrasov]. There, the *p*-band holes are in quasi-equilibrium: They are supplied by the pumping and removed via non-radiative recombination dominated by the holes leaking to the *d*-band. **The Fe vacancies control the density of the quasi-equilibrium *p*-holes. The hole relaxation is slowed down by absence of the Fe *d*- orbitals (missing pathways though which holes drift to top of the *d*- band).**

(2) Fig. 6 provides pictorial view for density of quasi-equilibrium *p*-type holes depend on the local composition. In the Fe-vacancy free [NQR_Texier] "true" metallic/ SC states ($K_{0.8}Fe_{2-x}Se_2$ x→0), the *p*-type hole density tends to zero and Cooper pair luminescence is suppressed. Since *p*-type holes are located ~1.3 eV below Fermi energy, they are not affected by free carriers at the $E_f$. This helps enormously to gain appreciable number of *p*-type holes and Cooper pair condensate at the same location. In fact, based on the fact that luminescence intensity is monotonically raising function of laser power, number of Cooper pairs is always larger than number of *p*-type holes available, $N_{cp} > N^{act}_h$: The observed luminescence is the *p*-hole limited.
   How is high density of Cooper pair condensate achieved? Since sample encompasses three phases, it's sufficient to consider superconductor to semiconductor and superconductor to insulator interfaces. At $T > 6$ K (the minimal temperature accessed experimentally) there is finite density of electronic state $N_n$ in the *n*-type semiconductor forming the active layers, Fig. 6. Classical semiconductors suffer from freeze-out at low temperatures. Degenerate semiconductors contain an impurity band which overlaps with the conduction or valence band, resulting in finite density of three carriers even at zero Kelvin. Alternatively, semiconductor may be a "bad metal" [Emery] due to real space electronic inhomogeneities and/or electronic correlations. In the clean limit, the penetration depth is estimated as $\xi_n$ ~ 1 nm (refer to the Supplementary Note #2), implying that Cooper pairs are available only in a narrow interface adjoining the SC phase, insert to Fig.
6. As for the superconductor to insulator interface, absence of free carriers
and in-gap levels inside the isolator result in negligible  penetration of
Cooper pairs into it. The pair potential Δ(x) → 0 at the boundary x=0 of the
SC with ideal insulator [Emig].  Since $N_{cp}$ → 0, the photoluminescence of
Cooper-pairs would be suppressed and would be a Cooper pair limited that was
not observed experimentally.

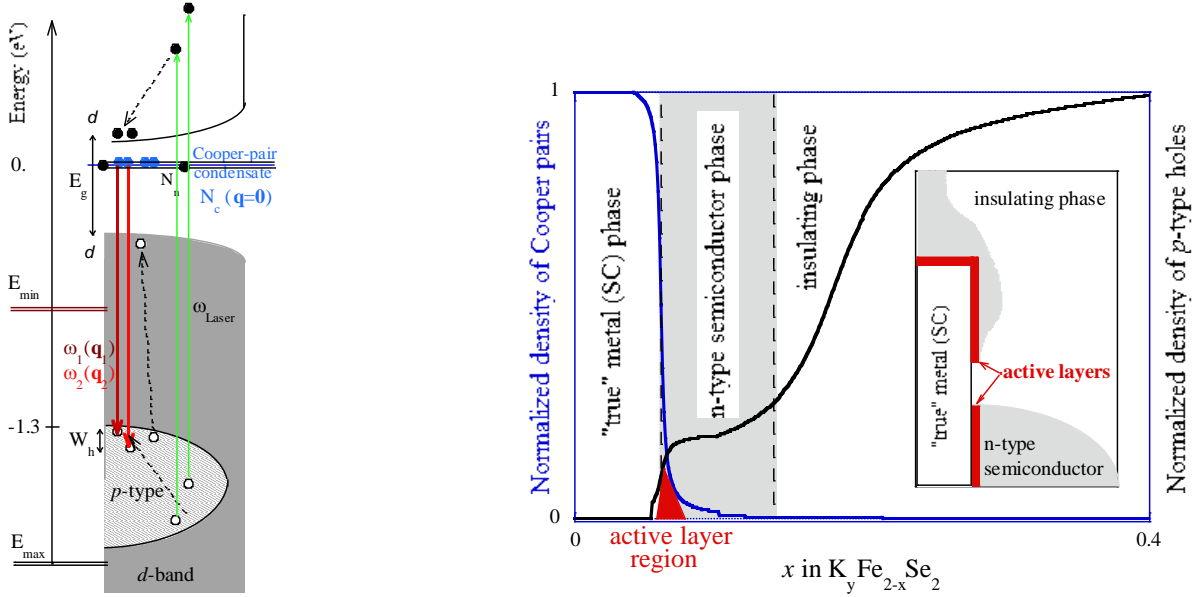

**Figure 5 (left). Schematic energy structure of degenerately doped semiconductor $K_yFe_{2-x}Se_2$ forming the active layer.** At $T > 6$ K (minimal temperature accessed in this study) there is always a finite density of electronic states at Fermi level, $N_n$. From luminescence vs. power dependence, FIG. 4, the luminescence is *p*-type hole limited in the whole *T*-range: there are more Cooper pairs available to recombine than *p*-type holes, and, therefore there is a sufficient density of electronic states out of which the Cooper pairs are formed by proximity effect [de Gennes] and/or by tunneling. Laser excitation (green arrows) couples occupied *p*- to empty *d*- via dipole transition for all allowed states in the BZ. Created *p*-type holes (open circles) relax to the top of the *p*-band and further to the top of the top of the *d*-band (dashed arrows). The entire Cooper pair (blue filled circles) recombine with pair of *E*-dispersed *p*-type holes, generating two photons (red arrows). The *E*-range of accessed luminescence is between $E_{min}$ = 0.95 and $E_{max}$ =1.8 eV, refer to Fig. 2.

**Figure 6 (right). Density of Cooper pairs and quasi-equilibrium *p*-type holes, depends on composition.** Cooper pair luminescence tends to avoid both "true" metallic phase $x \to 0$ [NQR_Texier] because number of *p*-type holes $N_h \to 0$ and insulating phase in which Cooper pairs do not penetrate to, $N_{cp} \to 0$. The composition boundaries of the *n*-type doped semiconducting phase (grey region) are presently not established well, but the phase presence is documented [ARPES_Chen, ND_Zhao]. Note that small fraction of the semiconducting phase (degenerately doped *n*-type semiconductor, Fig. 5) adjoined to the "true" metallic/SC phase forms the active layers (red area). Yet, the range of compositions (footprint of the red area)  allows for large number of SC/semiconducting interfaces to put up the network of active layers contributing to high luminescence output at the expense of broadened energy distribution of the *p*-type holes.
**Insert:** Schematic "real space" image. The Cooper pair luminescence takes place in a narrow layer of semiconductor phase, less than $\xi_n$ ~ 1 nm thick, that is adjoined to the "true" metallic/SC phase- it's essentially the interfacial phenomena. Conservative (low limit) estimate for number of active interfaces comes to ~$10^9$ per cm$^3$ (see text for details).

(3) Within the active layer in the semiconducting phase, Fig. 6, the Cooper
pair condensate overlaps with *p*-type quasi equilibrium holes. This opens the
door to radiative recombination studied theoretically by Asano *et. al.*

[Asano]. They evaluated the emission spectra of phonon in *p-n* junction with a SC metal and provided luminescence intensity in the first and the second orders. The second order recombination is dominated by a **resonant term** where an entire electron Cooper pair {**k**↑, -**k**↓} recombines with a pair of holes {-**k+q**↑, **k-q**↓} giving rise to a entangled photons pair (**q**,-**q**), shown in FIG. 2(a) in Ref. [Asano]. If pair were first to annihilate followed by one quasi particle (electron) recombining with the *p*-hole while the other electron recoiling to the continuum the photo luminesence integrated over broad enough energy range about the luminescence maximum shall be temperature independent because total number of electrons near Fermi energy is conserved. As it clearly seen in Fig. 2(a), the observed luminescence does not develop at the expense of the nearby background, providing a supporting evidence for predominant second order recombination. Though the model captures essential physics of single Cooper pair radiative recombination, a few experimental results need to be understood:

(a) The observed luminescence intensity vs. temperature dependence, depicted in FIG. 3, best fit as $I(T) \sim \Delta^2(T)/T$ for $0.6 < T/T_c < 1.0$ and $I(T) \sim \Delta^*(T)$ at $0.21 < T/T_c < 0.5$;

(b) The model relies on assumption that holes (i) have the same well-defined energy, though it doesn't elaborate on hole coherence and (ii) are always available to recombine with electron Cooper pairs. Since the luminescence is the *p*-type hole limited, the second condition is not met and/or holes are substantially less coherent than Cooper pair electrons;

(c) The model was developed to describe recombination of single Cooper pair, while in the KFS heterostructure one deals with an ensemble of radiating Cooper pairs, undoubtedly comprising **a partially coherent light source** characterized by polar diagram of radiant intensity, $I(\mathbf{R},\theta) \sim \cos^2(\theta) A(\mathbf{s},\mathbf{s},\omega)$, where $A(\mathbf{s},\mathbf{s},\omega)$ is the angular self-correlation function of the field in the far zone, and **s** is a unit vector in the **R** direction [Marchand&Wolf]. Due to experimental limitations (refer to Methods) angular distribution and coherent factors of the emitted radiation were not measured in this work.

Clearly, a model providing self-consistent description of both Cooper pair recombination and partly coherent field would be desirable to advance understanding and utilization of the KFS-like technology.

**Estimation of Internal Quantum Efficiency (IQE) of the KFS heterostructure.**

(1) In simplest microscopic model of the KFS heterostructure, the active layer is located at distance *x* inside the effective medium characterized by refractive index $n_{in}=1.821$, transmission $T_{in}=0.70$, and liner absorption $\alpha_{in}=0.031$ 1/nm for the incident laser light of $\lambda= 647.1$ nm and by $n_{out}=2.603$, $T_{out}=0.66$, and $\alpha_{out}=0.023$ 1/nm for the induced luminescence distributed around $\lambda \sim 930$ nm [Supplementary Note #3].

(2) Considering, for simplicity, that radiance vs. omega is a step-like function of the solid angle, R(Ω): R=*const* for Ω≤Ω' and R=0 for Ω'<Ω<2π, the propagation of the radiance is as follows:

$$R^0_{Las} \times [\Omega^0_{Las}/\Omega^*] \times T_{in} \, T_{out} \, exp\{-(\alpha_{in}+\alpha_{out})x\} \times (647/930)\eta \times [\Omega^0_{Spec}/\Omega^*] \times [\Omega^*/\Omega^0_{air}]^2 = R^0_{exp}$$

where $R^0_{Las}$ and $R^0_{exp}$ are the radiances of laser and the surface of the effective medium in the air; η is the IQE, defined as number of photons out (λ~ 930 nm) to number of photons in (λ= 647 nm), the 647/930 pre-factor accounts for energy loss if 1:1 conversion takes place; $\Omega^0_{Las}$ and $\Omega^0_{air}$ are solid angles of laser beam and radiation coming out of KFS medium, $\Omega^0_{Spec}$ is the solid angle of radiation admitted to the spectrometer, and $\Omega^*$ is solid angle in which luminescence is distributed inside the medium, refer to the insert to Fig. 7.

(3) There are three independent parameters in the above equation: $\Omega^*$, $x$, and η. To estimate *minimal value of* η we shall assume that (i) the active layer is located very close to the surface, $x \rightarrow 0$ and (ii) $\Omega^* \sim \Omega^0_{Spec}$. Recalling the numerical values of the parameters involved $R^0_{Las}$= 5.1x10$^6$ mW/(cm$^2$ Sr), $\Omega^0_{Las}$= 5.3x10$^{-3}$ Sr, $\Omega^0_{Spec}$= 1.9x10$^{-2}$ Sr, and $R^0_{exp}$= 2.86 x10$^3$ mW/(cm$^2$ Sr) one obtains η~ 0.29 for 10 K data [black solid curve in Fig. 2(a)]. The IQE temperature dependence follows the functional *T*- dependence depicted in Fig. 3, since it's directly proportional to the amplitude of radiance. Though our measurements were done on limited number of spots and may not be fully optimized to deliver highest possible IQE, the estimated value **rivals/exceed the IQE of the best human-made heterostructures** [PL_Suemune, R_Suemune]. Moreover, if less than 0.12 mW of laser power were used (Fig. 4) the "projected" IQE could reach 0.29x1.5= 0.44.

(4) The origin of large IQE deserves special consideration: The *p*-type holes generated in the Fe-vacancy-reach regions defuse into the active layers (Fig. 6), providing a higher local *p*-type holes concentration than that anticipated due to local laser light absorption. Since the luminescence is the *p*-type hole limited, the diffusion driven enlargement of the *p*- holes permits more electronic Cooper pairs to radiatively recombine with the holes, boosting the luminescence output. Due to high concentration of the SC/n-type semiconducting/AFM insulating heterostructures (> 10$^9$ cm$^{-3}$) KFS likely manifests a very efficient apparatus for harvesting the *p*-type holes. Under favorable supply rate of the *p*-holes, the local IQE of particular active layer could become larger than 1. Maximal value of η is limited by non-local energy conservation law that needs to be solved along with the radiance propagation equation.

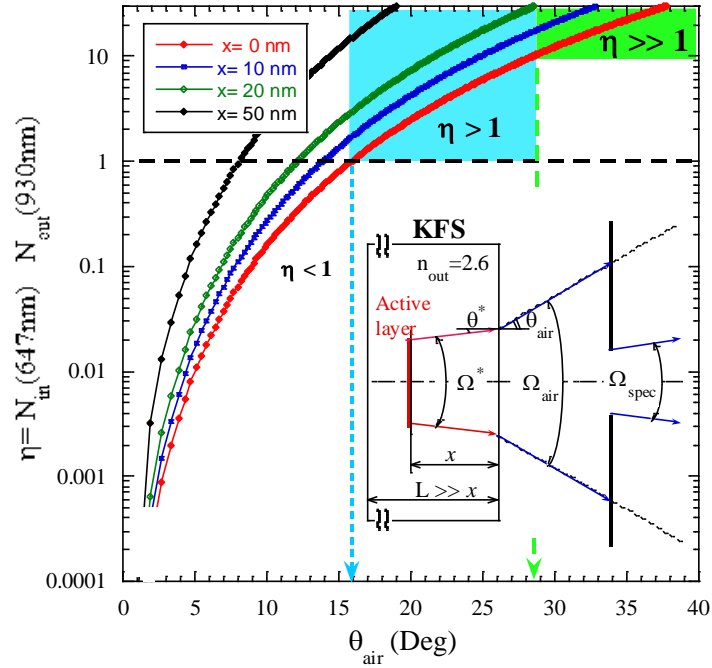

**Figure 7**. **Internal quantum efficiency, η (logarithmic scale) as a function of the $\theta_{air}$** - the angle that the emitted radiation forms with the normal to the surface, obtained from numerical solution of the radiance propagation equation. Red, blue, green, and black curves correspond to four locations of the active layer inside the KFS effective medium $x$=0, 10, 20, and 50 nm, respectively. For *minimal* η ($x$=0),  η becomes larger than 1.0 for $\theta_{air}$ > 15.9° (blue), and η > 10. for $\theta_{air}$ > 28.5° (green area). **Insert:** Emission of the KFS to the air (chamber of flow He cryostat). In the model adopted in the paper, the active layer is located at distance $x$ inside the effective medium, KFS.   $\Omega^0_{air} = 2\pi[1-\cos(\theta_{air})]$, $\Omega^* = 2\pi[1-\cos(\theta^*)]$, and $n_{out} \sin(\theta^*) = \sin(\theta_{air})$. $\Omega_{spec}$ is limited by aperture of the spectrometer.

(5) Numerical solution of the radiance propagation equation depicted in Fig. 7 as a set of 4 curves, representing IQE as function of the angle ($\theta_{air}$) that the emitted radiation forms with a normal to the surface for 4 locations of the active layer. If we *impose* the η ≤ 1 limit (horizontal dashed line), the entire emitted radiation is spread in a narrow solid angle, less than 2π[1-cos(θ=15.9°)]~0.24 Sr. Therefore, it should be relatively easy to verify experimentally whether the promising η > 1 case is realized in the KFS by collecting accurate angular distribution of the emitted radiance.

### KFS heretostructure is a super bright light emitter (SBLE) and EPPs source

(1) Comparing light emitting characteristics of the KSF heterostructure with those of typical semiconductor LED, one would realize that the KSF heterostructure has: (a) Higher internal quantum efficiency that makes it an attractive source for energy saving technological applications; and
(b) Higher spectral brightness [related to the spectral radiance shown in Fig. 2 via SB°(θ,λ)= SR°(θ,λ)/(hc/λ)], due to larger number of phonons emitted and smaller divergence. Combination of (a) and (b) in one unit constituents

KFS as a Super Bright Light Emitter (SBLE)- the prospective light source positioned between conventional LEDs and lasers.

(2) Theoretical studies suggested that SC/semiconducting heterostructure could be a source of the entangled photon pairs (EPPs) [Benson, Gywat], so much needed for quantum information processing and communication. However, EPPs generation has not yet been demonstrated experimentally [EL_Hayashi, PL_Hayashi, PL_Suemune, R_Suemune]. We suggest the following recipe for KFS-based EPPs source. It's critical to cleave off and use very thin chip of the KFS (L ~ 100-200 nm thick at the operational spot, refer to insert to FIG. 7). A "+q" group of photons escape through the right boundary of the chip propagating along the +X-axis, while the entangled counterparts(photons) propagate along the -X. Contrary to the SBLE (were L >> $x$) the entangled photons are not lost in the absorbing medium, but escape through the left boundary of the chip, forming "-q" group of photons. Position of the active layer(s) inside the chip, properties of semiconducting layers, left/right boundaries all contribute to a likely disparity between the emitted "+q" and "-q" photon beams.

(3) It's worth learning secrets that nature endowed in KFS, to replicate the technology in the similar human-designed units or to perfect the KFS-preparation technology. In this paper we demonstrated "proof-of-principles" operation of KFS as the SBLE. Replacing LEDs by SBLEs will likely go along with the progress in perfecting the technology. On the contrary, interest in utilizing KFS as the EPPs source may surge "as soon as it works".

**Cooper pair luminescence in high-energy electronic Raman scattering of high $T_c$ superconductors**

Local inhomogeneity is common properties of high $T_c$ superconductors, where the local metallic(SC) regions are often embedded in the insulating matrix. Atomic vacancies and substitution defects traps non-equilibrium light-generated carriers which can diffuse into the active layers where they radiactively recombine with superconducting condensate "leaking" from true metallic(SC)region. Not aiming to provide an overview, reported "rearrangement of spectral weight" (in $YBa_2Cu_3O_{6+x}$ [Rubhausen_1]) or "amplitude enhancement" (in $Bi_2Sr_2CaCu_2O_{8+\delta}$ [Rubhausen_2] and $HgBa_2CuO_{4+\delta}$ [PRL_Li]) of high-energy peaks (located at ~1500-2000 cm$^{-1}$ Raman shifts) upon cooling below $T_c$ could derive, in part, from luminescence of Cooper pairs.

**Conclusions**

To summarize, in $K_{0.75}Fe_{1.75}Se_2$ nature offers an elegant solution of the challenging technological problem of combining superconductor and semiconductors in nanostructure junctions: Superconducting Fe-vacancy free domains ($K_{y1}Fe_2Se_2$) are epitaxially bonded to semiconducting layers ($K_{y2}Fe_{2-x}Se_2$) characterized by Fe-vacancy gradient. The thin $n$-type semiconducting layers (~1 nm) adjoined the SC domains become the active layers. The electronic structure of the active layer, shown in Fig. 5, is tuned to induce Cooper pairs by proximity effect [de Gennes] and/or to accept Cooper pair tunneling from the SC phase. The semiconducting phase allows diffusion/injection of $p$-type holes generated in Fe-deficient regions by laser pumping. In the active regions of KFS, an ensemble of electron Cooper pair radiatively recombines with $p$-type holes, giving rise to partially coherent luminescence distributed about $\lambda$=930 nm. The internal quantum efficiency (IQE) of the laser-pumped KFS heterostructure, defined as $\eta$= $N_{out}(\lambda=930nm)/N_{in}(\lambda=647nm)$ (number of photons "out" to number of photons "in") is estimated to be close to 30%, rivaling or exceeding the IQE of the best human-made heterostructures [PL_Suemune, R_Suemune]. We demonstrated "proof-of-principles" of using KFS as a super bright light emitter (SBLE) and suggested an apparatus generating entangled photon pairs (EPPs).

## Methods

The crystal of iron-chalcogenide superconductors were grown by a self-melting method with nominal concentration of 0.8:2.1:2.0 (K:Fe:Se). The actual chemical composition was determined by EDXS as $K_{0.75}Fe_{1.75}Se_2$ (KFS). A major superconducting transition with $T_c^{middle}$ ~ 28 K was seen in resistivity curve. Further details of sample characterization can be found elsewhere [IR_Yuan].

We measured secondary emission, along the normal to the surface of the sample, out of five different spot positions produced by repeated cleaving of the same KFS sample: one spot, labeled M1, in the end of 2011 and four spots (M2, S1, S2, and S3) in the beginning of 2013. The experimental set up was based on standard high Raman shift layout and was not suitable for systematic mapping of angular distribution of luminescence intensity. The emission spectra were excited with Kr+ laser line of $\lambda$=647, 568, 413 nm with 0.04 to 9.7 mW of power focused into a spot of ~40x80 $\mu m^2$ on the freshly cleaved *ab*-plain crystal surface. The emitted light collected close to the backscattered geometry was focused to entrance slits of a spectrometer equipped with 150 lines/mm gratings. Princeton Instrument Si CCD and both the Si CCD and InGaAs linear detector were used in 2011 and 2013, respectively. The laser excitation was circularly polarized. In 2013 the polarization optics in the light collecting pathway was removed. Both detectors were calibrated using black body radiation standard (LabSphere) with known spectral radiance. Sample was mounted on a cold fingers of the He-flow cryostat (Oxford Instruments) cryostat providing *T*- stability better than 0.2 K. An estimated local heating in the laser spot did not exceed 4 K for laser power less than 1.5 mW. Typical data collection time was 30 or 60 seconds per scan with Si CCD and 300 seconds per scan with InGaAs detector.

## Acknowledgements

Authors thank Prof. Girsh Blumberg for helpful discussion and for making high-Raman shift setup available for this experiment. We are grateful to Prof. Ikuo Suemune and Prof. Yasuhiro Asano for stimulating discussion. A.I. acknowledge support by the US NSF-DMR-1104884. R.H.Y. and N.L.W. are supported by the NSFC and 973 projects of MOST, Grants No. 2011CB921701 and No.2012CB821403.

## Author contributions

A.I. planned high Raman shift experiments. A.I. carried out the experiment. R.H.Y. grew $K_yFe_{2-x}Se_2$ crystals and carried out resistivity, magnetization, specific heat, and optical reflectivity measurements, planned and coordinated by N.L.W. Luminescence data analysis was done by A.I. The paper was written by A.I. with input from N.L.W. All co-authors provided comments on the paper.

**Supplementary Note 1. Microscopic phase separation in $K_{0.75}Fe_{1.75}S_2$**

$K_{0.75}Fe_{1.75}Se_2$ (KFS) is an intrinsically inhomogeneous material [mic_Charnukha, TEM_Wang, IR_Yuan, STM_Li, IR_Homes, tr1_Ryan, NQR_Texier, ARPES_Chen, ND_Zhao], belonging to $A_yFe_{2-x}Se_2$ (A=Rb,K,Cs) family of Fe-chalcogenides. AFM occurs in the Fe-deficient lattice with Fe vacancies forming √5x√5x1 Fe vacancy-ordering structure. The parent AFM, $K_{0.8}Fe_{1.6}Se_2$ (245-structure), is an insulator with gap of ~0.5 eV [STM_Li, IR_Yuan]. The AFM phase occupies from ~50% from optical microscopy [mic_Charnukha], to as high as ~96% from NQR [NQR_Texier]), though the majority of experiments indicated ~85-90% fractional occupancy [IR_Homes, tr1_Ryan]. Stoichiometry and morphology of the metallic (SC at low T) phase remains less settled. STM [STM_Li] and NQR [NQR_Texier] indicate that the SC phase does not contain Fe vacancies or magnetic moments. TEM suggests that SC could have the Fe-vacancy disordered states [TEM_Wang].

Experimental techniques have different probing length scales and, therefore, different sensitivity to the Fe-vacancy concentration. Scattering in the reported amount of the metallic phase may be understood assuming that phase separation encompass three phases: (i) "true" metallic (SC) phase that does not contain Fe vacancies or magnetic moments, (ii) Fe-vacancy ordered insulating AFM phase, and (iii) partly Fe-vacancy ordered, intermediate semiconducting phase. The latter serve as a buffer between SC and AFM phases where the gradient in the Fe-vacancy concentration is leveled out. From structural perspective, the intermediate semiconducting phase may look like nanoseparated vacancy-disordered presumably metallic sheets in the bulk with an unknown in-plane form factor [TEM_Wang]. ARPES [ARPES_Chen] and transport measurements in Refs [ARPES_Chen, ND_Zhao] have identified several different phases in the $K_yFe_{2-x}Se_2$, namely a superconducting phase with only electron-like Fermi surface [ARPES_Chen], an insulating phase with about ~0.5 eV band gap, and a small gap semiconducting phase.

**Supplementary Note 2. Estimate for Cooper pair penetration depth into the n-type semiconductor**

The pair potential Δ(x) at the interface of superconductor (x<0) to normal metal (x≥0), assuming no external field, $T \sim T_c$, and no attractive or repulsive "self" interactions outside the SC can be written as [de Gennes]:

$$\Delta(x) \propto \Delta_0 \frac{N_n}{N_{sc}} W \exp(-\frac{x}{\xi_n})$$

where $\Delta_0$ is the pair potential in the bulk of superconductor; $N_n$ and $N_{sc}$ are the density at Fermi levels in normal metal and superconductor (above $T_c$), W is transmission coefficient, and $\xi_n = \frac{\hbar v_n}{k_B T} = \frac{\hbar^2 (3\pi^2 n)^{1/3}}{k_B T m^*}$. Here $v_n$ is the Fermi velocity, $m^*$ is the effective mass, and n in carriers concentration in the normal metal.

In the "dirty" limit, the equation for the pair potential remains the same, but $\xi_n = \sqrt{\dfrac{D}{2\pi T}}$ where $D$ is a diffusion constant $D = \dfrac{1}{3} v_n l_n$, $v_n$ is the Fermi velocity and $l_n$ is a mean free path in the "dirty" metal.

The electron concentration in SC domains in KFS is $10^{18}$ cm$^{-3}$ [IR_Homes] or higher. For the purpose of estimating $\xi_n$ (in the clear limit), we shall assume that (1) the electron concentration in $n$-type semiconductor epitaxially attached to the SC domain and having very few Fe vacancies is n=$10^{17}$ cm$^{-3}$ (ten times less than in the SC domain), and that (2) $T\sim$ 30 K, and (3) m*= 4 m$_e$ (typical mass renormalization for Fe-chalcogenides). We obtain $\xi_n \sim$ 1 nm showing that Cooper pairs are available in narrow layers adjoining the SC phase.

Finite pair potential $\Delta(x)$ in the active layer may also help $p$-type holes to accrue partial coherence. The mechanism deserve theoretical consideration that is beyond the scope of present study.

## Supplementary Note 3. Optical constants for KFS effective medium.

Based on near-field optical microscopy [mic_Charnukha] and TEM studies [TEM_Wang] the KFS may be viewed as self assembled cells structure: facets of individual cell/parallelepiped are formed by SC and $n$-type semiconducting phases, while the cell's core is build of the insulating AFM phase. To estimate internal quantum efficiency of the KFS heterostructure we adopted a simple model assuming that a single active layer is located at distance $x$ inside an effective KFS medium characterized by effective optical constants: the refractive index $n(\omega)$ and the extinction coefficient $k(\omega)$ shown in the plot below.

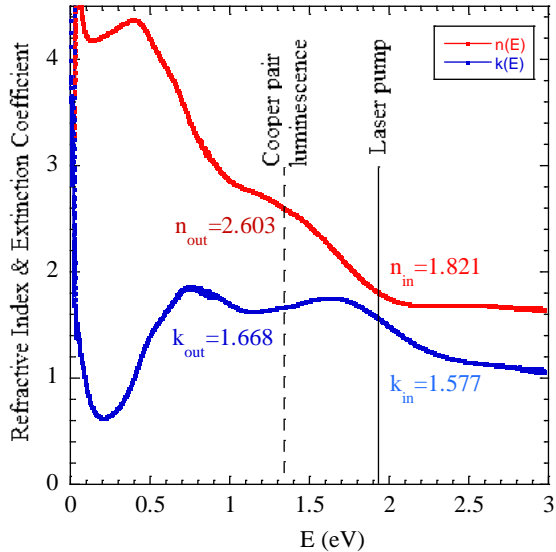

The transmission, $T$(E) and linear absorption, $\alpha$(E) are given by the following equations:

$T(E) = 4n(E) / [(n(E)+1)^2 + k^2(E)]$
$\alpha(E) = 4\pi k(E) / \lambda$

where $\lambda$ is the wavelength in vacuum. Using numerical values of refractive indexes and extinction coefficient shown in the plot, one obtains T$_{in}$=0.70 and $\alpha_{in}$=0.031 1/nm for the incident laser light of $\lambda$= 647.1 nm and T$_{out}$=0.66, and $\alpha_{out}$=0.023 1/nm for the luminescence distributed around $\lambda\sim$ 930 nm.

**FIG. S1.** Refractive index (red) and extinction coefficient (blue) as function of energy, obtained from dielectric constants $\varepsilon_1(\omega)$ and $\varepsilon_2(\omega)$ derived from optical reflectivity measurements on K$_{0.75}$Fe$_{1.75}$Se$_2$ sample from the same batch at 300 K [IR_Yuan]. In the $E$-range of interest, the dielectric constants exhibit week temperature dependence.

**References**

[van Wees] B. J. van Wees and H. Takayanagi, in Mesoscopic Electron Transport, NATO ASI Series E Vol. **345** (Kluwer Academic Publishers, Dordrecht, 1997), p. 469.

[EL_Hayashi] Y. Hayashi, K. Tanaka, T. Akazaki, M. Jo, H. Kumano, and I. Suemune. Superconductor-based Light Emitting Diode: Demonstration of Role of Cooper Pairs in Radiative Recombination Processes. Appl. Phys. Express **1**, 011701 (2008).

[PL_Hayashi] Y. Hayashi, M. Jo, H. Kumano, and I. Suemune, Luminescence of n-InGaAs/p-InP Light Emitting Diode with Superconductiong Nb Electrode, Report.

[PL_Suemune] I. Suemune, Y. Hayashi, S. Kuramitsu, K. Tanaka, T. Akazaki, H. Sasakura, R. Inoue, H. Takayanagi, Y. Asano, E. Hanamura, S. Odashima, and H. Kumano. A Cooper-Pair Light-Emitting Diode: Temperature Dependence of Both Quantum Efficiency and Radiative Recombination Lifetime. Applied Physics Express **3**, 054001 (2010).

[R_Suemune] I. Suemune, H. Sasakura, Y. Hayashi, K. Tanaka, T. Akazaki, Y. Asano, R. Inoue, H. Takayanagi, E. Hanamura, J. Huh, C. Hermannstädter, S. Odashima, and H. Kumano. Cooper-Pair Radiative Recombination in Semiconductor Heterostructures: Impact on Quantum Optics and Optoelectronics. Japanese Journal of Applied Physics **51**, 010114 (2012).

[Benson] O. Benson, C. Santori, M. Pelton, and Y. Yamamoto. Regulated and Entangled Photons from a Single Quantum Dot. Phys. Rev. Lett. **84**, 2513 (2000).

[Gywat] O. Gywat, G. Burkard, and D. Loss. Biexcitons in coupled quantum dots as a source of entangled photons. Phys. Rev. B **65**, 205329 (2002).

[mic_Charnukha] A. Charnukha, A. Cvitkovic, T. Prokscha, D. Pröpper, N. Ocelic, A. Suter, Z. Salman, E. Morenzoni, J. Deisenhofer, V. Tsurkan, A. Loidl, B. Keimer, and A. V. Boris. Nanoscale Layering of Antiferromagnetic and Superconducting Phases in $Rb_2Fe_4Se_5$ Single Crystals. Phys. Rev. Lett. **109**, 017003 (2012).

[TEM_Wang] Z. Wang, Y. J. Song, H. L. Shi, Z.W. Wang, Z. Chen, H. F. Tian, G. F. Chen, J. G. Guo, H. X. Yang, and J. Q. Li. Microstructure and ordering of iron vacancies in the superconductor system $K_yFe_xSe_2$ as seen via transmission electron microscopy. Phys. Rev. B **83**, 140505 (2011).

[de Gennes] P. G. de Gennes *Superconductivity of Metals and Alloys* (Benjamin, New York, 1969).

[Asano] Y. Asano, I. Suemune, H. Takayanagi, and E. Hanamura. Luminescence of a Cooper Pair. Phys. Rev. Lett. **103**, 187001 (2009).

[Hanamura] E. Hanamura. Superradance from p–n Junction of Hole- and Electron-Superconductors. Phys. Status Solidi B **234**, 166 (2002)

[IR_Yuan] R. H. Yuan, T. Dong, Y. J. Song, P. Zheng, G. F. Chen, J. P. Hu, J. Q. Li, and N. L. Wang. Nanoscale phase separation of antiferromagnetic order and superconductivity in $K_{0.75}Fe_{1.75}Se_2$. Sci. Rep. **2**, 221 (2012).

[STM_Li] W. Li, H. Ding, P. Deng, K. Chang, C. Song, K. He, L. Wang, X. Ma, J.-P. Hu, X. Chen, and Q.-K. Xue. Phase separation and magnetic order in K-doped iron selenide superconductor. Nat. Phys. **8**, 126 (2012).

[IR_Homes] C. C. Homes, Z. J. Xu, J. S. Wen, and G. D. Gu. Effective medium approximation and the complex optical properties of the inhomogeneous superconductor $K_{0.8}Fe_{2-y}Se_2$. Phys. Rev. B **86**, 144530 (2012).

[tr1_Ryan] D. H. Ryan, W. N. Rowan-Weetaluktuk, J. M. Cadogan, R. Hu, W. E. Straszheim, S. L. Bud'ko, and P. C. Canfield. $^{57}$Fe Mössbauer study of magnetic ordering in superconducting $K_{0.80}Fe_{1.76}Se_{2.00}$ single crystals. Phys. Rev. B **83**, 104526 (2011).

[NQR_Texier] Y. Texier, J. Deisenhofer, V. Tsurkan, A. Loidl, D. S. Inosov, G. Friemel, and J. Bobroff. NMR Study in the Iron-Selenide $Rb_{0.74}Fe_{1.6}Se_2$: Determination of the Superconducting Phase as Iron Vacancy-Free $Rb_{0.3}Fe_2Se_2$ . Phys. Rev. Lett. **108**, 237002 (2012).

[ARPES_Chen] F. Chen, M. Xu, Q.Q. Ge, Y. Zhang, Z.R. Ye, L.X. Yang, J. Jiang, B.P. Xie, R.C. Che, M. Zhang, A.F. Wang, X.H. Chen, D.W. Shen, J.P. Hu, and D.L. Feng. Electronic Identification of the Parental Phases and Mesoscopic Phase Separation of $K_xFe_{2-y}Se_2$ Superconductors. Phys. Rev. X **1**, 021020 (2011).

[ND_Zhao] J. Zhao, H. Cao, E. Bourret-Courchesne, D.-H. Lee, and R. J. Birgeneau. Neutron-Diffraction Measurements of an Antiferromagnetic Semiconducting Phase in the Vicinity of the High-Temperature Superconducting State of $K_xFe_{2-y}Se_2$. Phys. Rev. Lett. **109**, 267003 (2012).

[LDA_Nekrasov] I. A. Nekrasov and M. V. Sadovskii. Electronic structure, topological phase transitions and superconductivity in $(K, Cs)_xFe_2Se_2$. JETP Letters, **93** 166 (2011).

[Emery] V. J. Emery and S. A. Kivelson. Superconductivity in Bad Metals. Phys. Rev. Lett. **74**, 3253 (1995).

[Emig] T. Emig, K. Samokhin, and S. Scheidl. Charge modulation at the surface of high-$T_c$ superconductors. Phys. Rev. B **56** 8386, (1997).

[Marchand&Wolf] E.W. Marchand and E. Wolf. Radiometry with sources of any state of coherence. J. of Opt. Soc. of Am., **64** 1219 (1974).

[Rubhausen_1] M. Rübhausen, C. T. Rieck, N. Dieckmann, K.-O. Subke, A. Bock, and U. Merkt. Inelastic light scattering at high-energy excitations in doped cuprate superconductors. Phys. Rev. B **56** 14797, (1997).

[Rubhausen_2] M. Rübhausen, O. A. Hammerstein, A. Bock, U. Merkt, C. T. Rieck, P. Guptasarma, D. G. Hinks, and M. V. Klein. Doping Dependence of the Electronic Interactions in Bi-2212 Cuprate Superconductors: Doped Antiferromagnets or Antiferromagnetic Fermi Liquids? Phys. Rev. Lett. **82** 5349 (1999).

[PRL_Li] Y. Li, M. Le Tacon, M. Bakr, D. Terrade, D. Manske, R. Hackl, L. Ji, M. K. Chan, N. Barisic, X. Zhao, M. Greven, and B. Keimer. Feedback Effect on High-Energy Magnetic Fluctuations in the Model High-Temperature Superconductor $HgBa_2CuO_{4+\delta}$ Observed by Electronic Raman Scattering. Phys. Rev. Lett. **108**, 227003 (2012).